\documentclass[twocolumn]{aastex631}

\usepackage{lineno}
\usepackage{microtype}
\usepackage{hyperref}
\usepackage{tabularx}
\usepackage{makecell}
\usepackage{footnote}
\usepackage{float}
\usepackage{footmisc}

\usepackage[utf8]{inputenc}

\begin{document}

\title{Discovery of Gaia17bpp, a Giant Star with the Deepest and Longest Known Dimming Event}

\author[0000-0003-0484-3331]{Anastasios Tzanidakis}\thanks{Corresponding Author: \href{mailto:atzanida@uw.edu}{atzanida@uw.edu}}
\affiliation{Department of Astronomy and the DiRAC Institute, University of Washington, 3910 15th Avenue NE, Seattle, WA 98195, USA}

\author[0000-0002-0637-835X]{James R. A. Davenport}
\affiliation{Department of Astronomy and the DiRAC Institute, University of Washington, 3910 15th Avenue NE, Seattle, WA 98195, USA}

\author[0000-0001-8018-5348]{Eric C. Bellm}
\affiliation{Department of Astronomy and the DiRAC Institute, University of Washington, 3910 15th Avenue NE, Seattle, WA 98195, USA}

\author[0000-0001-5538-0395]{Yuankun Wang}
\affiliation{Department of Astronomy and the DiRAC Institute, University of Washington, 3910 15th Avenue NE, Seattle, WA 98195, USA}

\begin{abstract}
We report the serendipitous discovery of Gaia17bpp/2MASS J19372316+1759029, a star with a deep single large-amplitude dimming event of $\sim$4.5 magnitudes that lasted over 6.5 years. Using the optical to IR spectral energy distribution (SED), we constrain the primary star to be a cool giant M0-III star with effective temperature $T_{\text{eff}}$=3,850 K and radius R=58 R$_{\odot}$. Based on the SED fitting, we obtained a bimodal posterior distribution of primary stellar masses at 1.5 M${\odot}$ and 3.7 M${\odot}$. Within the last 66 years of photometric coverage, no other significant dimming events of this depth and duration were identified in the optical light curves. Using a Gaussian Process, we fit a high-order Gaussian model to the optical and IR light curves and conclude the dimming event exhibits moderate asymmetries from optical to IR. At the minimum of the dimming event, the (W$_{1}$-W$_{2}$) color was bluer by $\sim$0.2 mag relative to the primary star outside the dimming event. The ingress and egress colors show a shallow reddening profile. We suggest that the main culprit of the dimming event is likely due to the presence of a large, optically thick disk transiting the primary giant star. By fitting a monochromatic transit model of an oblate disk transiting a star, we found good agreement with a slow-moving, 0.005 km sec$^{-1}$, disk with a $\sim$1.4 AU radius. We propose that Gaia17bpp belongs to a rare binary star population similar to the Epsilon Aurigae system, which consists of a secondary star enshrouded by an optically thick debris disk. 
\end{abstract}


\section{Introduction} \label{sec:Intro}

The long-term photometric monitoring of the sky is steadily increasing the discovery of remarkable dimming events on the timescale of decades. Amongst slowly evolving stellar variables, an emerging population of photometrically deep, long-duration, and long-period anomalous binary systems with the secondary star being enveloped by a large disk that calls for more attention \citep{2016SPIE.9907E..17S}. At the turn of the century, the enigmatic star Epsilon Aurigae ($\epsilon$-Aur), which exhibits a single flat-bottomed, $\sim$2 year long dimming event, challenged our understanding of binary stars \citep{1991ApJ...367..278C}. Today, more than a few decades later, $\epsilon$-Aur has been understood to be a low-mass F0 supergiant star surrounded by a young B5V-type star companion with an extended cool disk with an orbital period of 27 years. Both spectroscopic and direct interferometric imaging techniques have confirmed the presence of a geometrically thin and optically thick disk that is consistent with debris \citep{2011AJ....142..174S, 2010Natur.464..870K}. Open questions such as the origins and lifetime of the disk, occurrence rates, and formation channels of such systems remain unexplored. \citet{2018MNRAS.476.5026G} conducted a theoretical study using MESA modules to explain the evolution of the $\epsilon$-Aur binary system. Their study suggests the formation of the large debris disk found around the secondary star of $\epsilon$-Aur was the product of accretion from two intermediate-mass stars, 9.85 M$_{\odot}$, and 4.5 M$_{\odot}$, with an initial orbital period of roughly 100 days, after the primary star had made its first ascent to the post-RGB/pre-AGB phase. As noted, however, their models could not replicate the large orbital period of $\epsilon$-Aur which was likely due to poor constraints on the mass transfer profile. 

Prior to this work, TYC 2505-672-1 \citep{2016A&A...588A..90L} previously held the record with the longest duration dimming event and longest orbital period binary. The system consists of an M2-III giant that undergoes a 4.5 magnitude 3.5-year long deep eclipse every 69.1 years \citep{2016AJ....151..123R}. It is suggested that the companion star was a cool subdwarf-B (sdB) type star that will eventually become a low-mass pre-Helium white dwarf (Pre-He WD), enshrouded by an opaque disk. The origin of the disk, however, remains puzzling since accretion from the primary would be challenging to explain given its large semi-major axis separation \citep{2016AJ....151..123R}. Additionally, previous searches for Pre-He WD stars have been found in compact binary systems \citep{2018MNRAS.475.2560V, 2014MNRAS.437.1681M}, unlike a very long binary such as TYC 2505-672-1. More recently, reminiscent systems such as VVV-WIT-08 \citep{WIT08_ref}, a K7 giant star that showed a single smooth dimming event have also been thought to be occulted by an optically thick disk. Despite the limited understanding of the occulting disk in the majority of these dimming stars, a significant number of them exhibit similar characteristics. These include an evolved primary star, prolonged periods of single deep dimming events that last for years, large orbital periods, and the absence of IR excess. Finally, other identified long-period binaries have also claimed the presence of an extended disk around the companion star, such as ASASSN-21co \citep{2021RNAAS...5..147R} and Eta Gem \citep{2022MNRAS.516.2514T} that cause prolonged deep eclipses.

In this study, we report the serendipitous discovery of a new giant dimming star, Gaia17bpp/2MASS J19372316+175902, which now holds the record for the longest and deepest dimming event. We incorporate the use of multi-wavelength light curves and optical spectra of this star to understand the nature of the dimming mechanism. This paper is organized as follows. In Section \S \ref{sec:Data} we discuss the data acquisition from all-sky surveys and spectroscopic follow-up efforts. Section \ref{sec:PrimaryStar} describes the modeling of the spectral energy distribution (SED) of the primary star using multi-band archival photometry. Similarly, in Section \ref{sec:DimmingEventProperties} we analyze the properties and shapes of the light curves from the optical to infrared wavelengths. We provide an interpretation of the dimming event, fit a monochromatic oblate disk model using IR light curves, and investigate the possible mechanism driving it in Section \S \ref{sec:Discussion}. Finally, in Section \S \ref{sec:conclusion} we provide our conclusions.

\section{Data} \label{sec:Data}

In the following section, we describe the data acquisition from multiple public all-sky surveys. All collated time-series photometric measurements have been combined to generate a mosaic light curve of Gaia17bpp spanning from the late 1950s to the present date. 

\subsection{Gaia}  

\subsubsection{Gaia Photometric Science Alerts}

\begin{figure}
\includegraphics[width=0.45\textwidth]{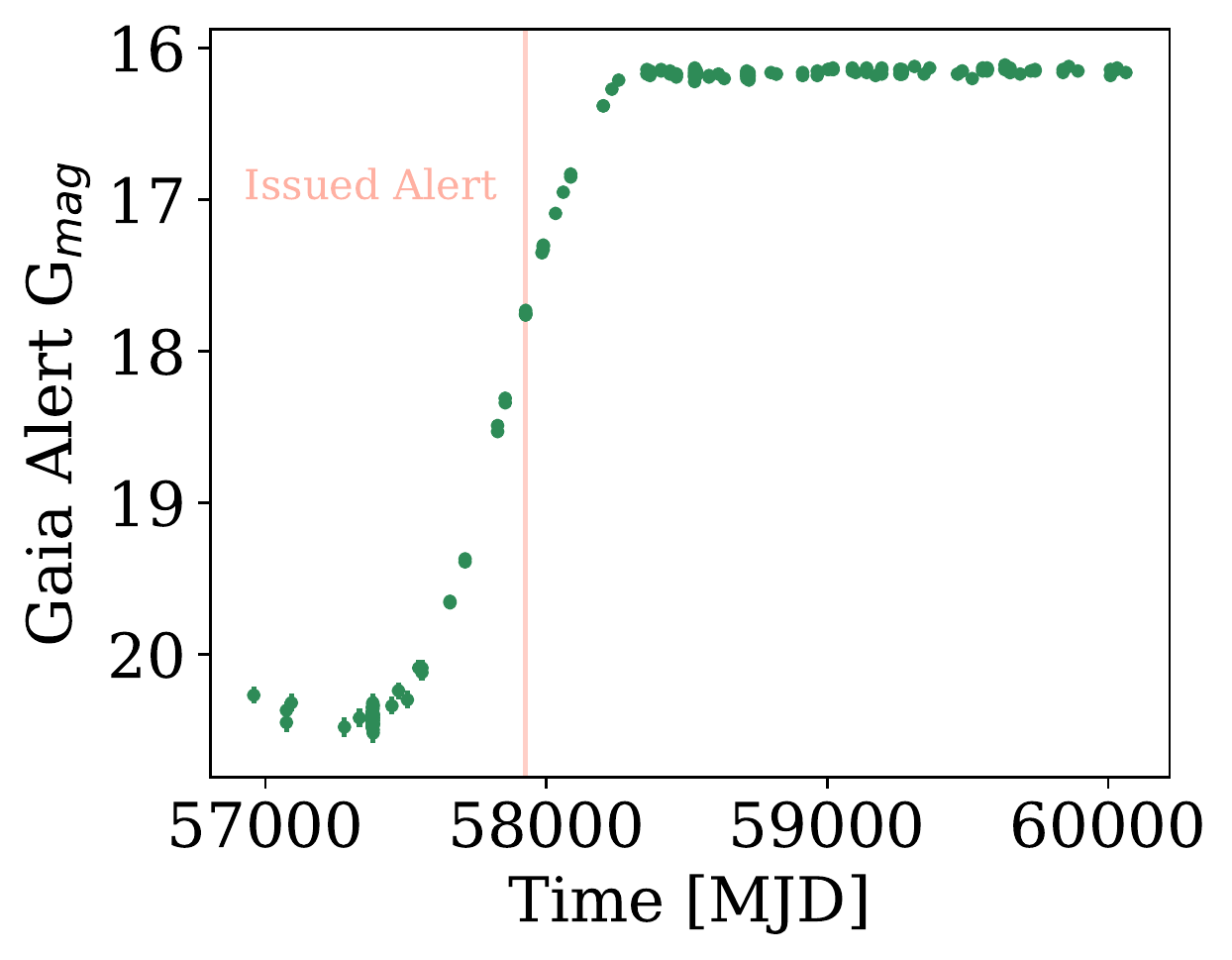}
\caption{Gaia17bpp issued alert light curve in Gaia $G$. The red line indicates the issued alert epoch.}
\label{fig:gaia_discovery_lc}
\end{figure}

Gaia17bpp was first saved as an alert through the Gaia photometric alert stream \citep{2017TNSTR.690....1D}, and was discovered through a subsequent search for deep and long ($>$5 years) stellar variability. Gaia has been scanning the night sky since 2014 and has been collecting photometric measurements in the Gaia $G$ bandpass and releasing alerts through the Gaia Photometric Science Alerts (GPSA) described in \cite{2021A&A...652A..76H}. We query the Gaia alert light curves through the Gaia Alert website portal\footnote{\href{http://gsaweb.ast.cam.ac.uk/alerts/home}{http://gsaweb.ast.cam.ac.uk/alerts/home}} to collect all the available epochal Gaia $G$, BP, and RP detections. In Figure \ref{fig:gaia_discovery_lc} we show the issued alert Gaia $G$ light curve of Gaia17bpp.

We note that the current alert schema does not include photometric uncertainties, however, \citet{2021A&A...652A..76H} suggests an empirical means of estimating uncertainties as a function of source brightness. We also mined\footnote{\href{https://github.com/AndyTza/GaiaAlertsPy}{https://github.com/AndyTza/GaiaAlertsPy}} the uncalibrated BP/RP flux spectra following the approach of \cite{2021A&A...652A..76H}. Using the integrated BP/RP spectra as a reasonable estimate of the color evolution in the optical regime. The BP photometry in this case suffers from higher scatter because the baseline BP photometry is close to the limiting magnitude. To examine the difference between the estimated uncalibrated BP and RP photometry we crossmatched all photometric alerts that were also found in the Gaia Data Release 3 \citep{gdr3} and compared the reported BP and RP magnitudes with the ones derived from this work. On average, we found a standard deviation of 1.5 mag per magnitude bin. For our reported BP/RP detections, we inflated the photometric errors to capture any excess uncertainty by 1.5 $\sigma_{G}$.\\

\break

\subsubsection{Gaia Data Releases}
According to the full photometric history enabled through the GPSA, Gaia17bpp was already at its faintest phase and close to the detection threshold of Gaia when observations began. We identified the system in both Gaia DR2 and DR3 catalogs and investigate the astrometric solutions, specifically the parallax and proper motion vectors. In both data releases the parallax is negative. We suspect this is due to the photometric uncertainty at the bottom of the dimming event, leading to a negative parallax solution \citep{2018A&A...616A...9L}. To obtain distances we rely on the photo-geometric distances derived from the work of \cite{2021AJ....161..147B} that compensates for negative parallax while enabling a strong prior from the available photometry and Galactic extinction at the line of sight. We adopt the median photo-geometric posterior distribution distance of 8.5$^{+2.3} _{-1.6}$ kpc throughout all calculations in this study. In the presence of negative parallaxes, we caution that this estimated distance is mostly prior dominated. In Table \ref{table:basic-info} we highlight all the key parameters reported from Gaia DR3.

\begin{figure}
  \centering
  \includegraphics[width=0.45\textwidth]{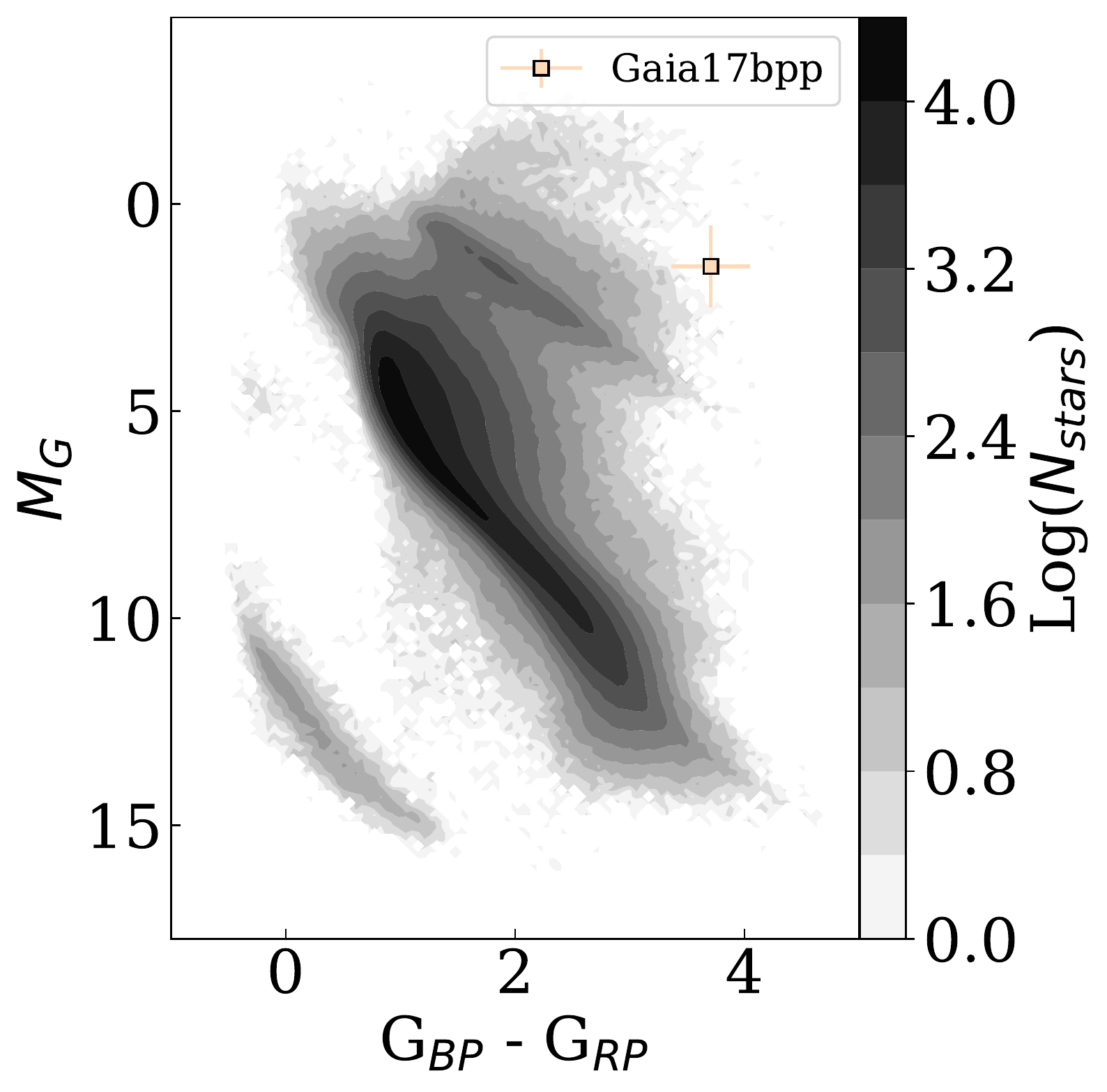}
  \caption{Color magnitude diagram (in gray) for the nearby Gaia DR3 sample \citep{gdr3} shaded by the number of stars per bin. We over-plot the position of Gaia17bpp on the CMD from the most recent BP/RP fluxes and Gaia-G estimates using GPSA.}
  \label{fig:gaia_cmd}
\end{figure}

Enabled through the distance estimate and extrapolation of BP/RP magnitudes, we placed the location of Gaia17bpp on the observed Hertzsprung-Russell (HR) diagram at the present date is seen in Figure \ref{fig:gaia_cmd}. We also attempted to track the evolution of Gaia17bpp throughout its dimming phase. We bin both the BP/RP and G band magnitudes by a 450-day running median to capture the overall color and absolute magnitude evolution. In general, we found consistency in the reported photometry with both the official Gaia data releases and the photometric science alerts. However, we noticed that the dimming event was predominantly bluer than the quiescent star. As noted by \cite{2021A&A...649A...3R}, faint sources with red colors tend to have overestimated BP fluxes. To mitigate the effects of overestimating the BP flux, we only computed the BP/RP magnitudes based on the epochal spectra after 2016 when the source began re-brightening.

\begin{table}[ht]
\centering
\begin{tabular}{@{}cc@{}}
\hline\hline
Survey Info & Parameters \\
[0.05ex]
\hline
Gaia ID & 1824311891830344704 \\
Coordinates ($\alpha_{J2016}$, $\delta_{J2016}$) & 19h37m23.16s, 17d59m2.90s \\
Parallax\footnote{Data source from Gaia DR3 \label{note}} ($\omega$) & -0.183 $\pm$ 0.403 mas \\
Distance\footnote{Photogeometric distance posterior distribution \cite{2021AJ....161..147B}} & 8462 $^{+2317}_{-1567}$ pc \\
Proper Motion$^{\ref{note}}$ RA ($\mu_{\alpha \cos(\delta)}$) & -0.238 $\pm$ 0.354 mas yr$^{-1}$ \\
Proper Motion$^{\ref{note}}$ DEC ($\mu_{\delta}$) & -3.046 $\pm$ 0.348 mas yr$^{-1}$ \\
\hline\hline
\end{tabular}
\caption{Astrometric solutions based on the Gaia DR3 catalog for Gaia17bpp. \label{table:basic-info}}
\end{table}

\subsection{WISE}
Subsequent investigation revealed that Gaia17bpp was also well monitored by the Wide-field Infrared Survey Explorer (WISE) \citep{wise_1, wise_2} and NEOWISE \citep{2011ApJ...731...53M} in the W$_{1}$-W$_{4}$ filters. To query the WISE and NEOWISE light curves we use an open-source Python tool, \texttt{WISE Light Curves}\footnote{\href{https://github.com/HC-Hwang/wise_light_curves}{https://github.com/HC-Hwang/wise$\_$light$\_$curves}} that automatically queries NASA/IPAC Infrared Science Archive to compile the epochal photometry from both surveys. We use a five-arcsecond cone search query and identify WISE J193723.16+175903 and confirm it is the identical source to Gaia17bpp. We also searched through the AKARI and Planck observations and found no excess flux within 1 deg$^{2}$ in the far-infrared.
\hfill \break

\subsection{MAST}
We also queried the Barbara A. Mikulski Archive for Space Telescopes (MAST) data archive and found a few optical observations from the Panoramic Survey Telescope and Rapid Response System (PS1; \cite{ps1}), two NUV images from the Galaxy Evolution Explorer (GALEX; \cite{galex_citation}) in the All-Sky Imaging Survey (AIS), and more recently three Full-Frame Images from the Transiting Exoplanet Survey Satellite (TESS; \cite{tess_citation}). We generated the PS1 light curves using the Space Telescope Science Institute database to query the PS1 Detection table fields\footnote{\href{https://catalogs.mast.stsci.edu/panstarrs/}{https://catalogs.mast.stsci.edu/panstarrs/}} with a crossmatch cone radius of 3 arcseconds. The PS1 detection table contains single epoch photometry from single visit exposure that produces the PSF aperture photometry of detected sources. Both GALEX NUV raw count images were visually inspected, however, no coincident source at the location of the Gaia17bpp system was found. We also searched the GALEX GR6/GR7 sources and only found the nearest source to be 1.4 arcminutes away, making it unlikely this is our target. Based on the GALEX AIS 20.5 mag NUV limiting magnitude, we conclude that Gaia17bpp must be fainter.

\subsection{Archival Photometry}
We obtained archival photometry using the Carlsberg Meridian Catalogue 15 (CMC15; \cite{1993BICDS..42....5F}), Initial Gaia Source List (IGSL; \cite{2014A&A...570A..87S}), USNO-B1 all-sky catalog (USNO-B; \cite{2003AJ....125..984M}), Guide Star Catalog (GSC; \cite{1994Ap&SS.217...31J}), Palomar Observatory Sky Survey (POSS; \cite{1991PASP..103..661R}), INT Photometric H-Alpha Survey (IPHAS; \cite{2004AAS...20511303W}), and Digital Access to a Sky Century at Harvard (DASCH; \cite{2017ASSP...50..203G}). The DASCH photographic plates while having been scanned had not reported observed magnitudes and extractions at the location of the star. Instead, we ran a source extraction tool to perform aperture photometry at the location of Gaia17bpp using \texttt{SEP} \citep{2018ascl.soft11004B}. For each 400x400 arcsecond scanned plate from DASCH, we crossmatched each 2$\sigma$ source detection to the PS1 r-band catalog assuming a limiting magnitude of 16 mag. From each science image, we subtract out the background root mean square variation. After comparing, and extrapolating the zero-point calibration between the PS1 and DASCH images, we summed the fixed aperture at the location of Gaia17bpp to extrapolate a rough estimate of its magnitude and error\footnote{Given the distance, proper motion, and coordinates obtained from Gaia DR3 during the J2016.0 epoch, we estimate the change in position due to proper motion to be approximately 0.2 arcseconds.}.

\begin{figure*}
\includegraphics[width=0.93\textwidth]{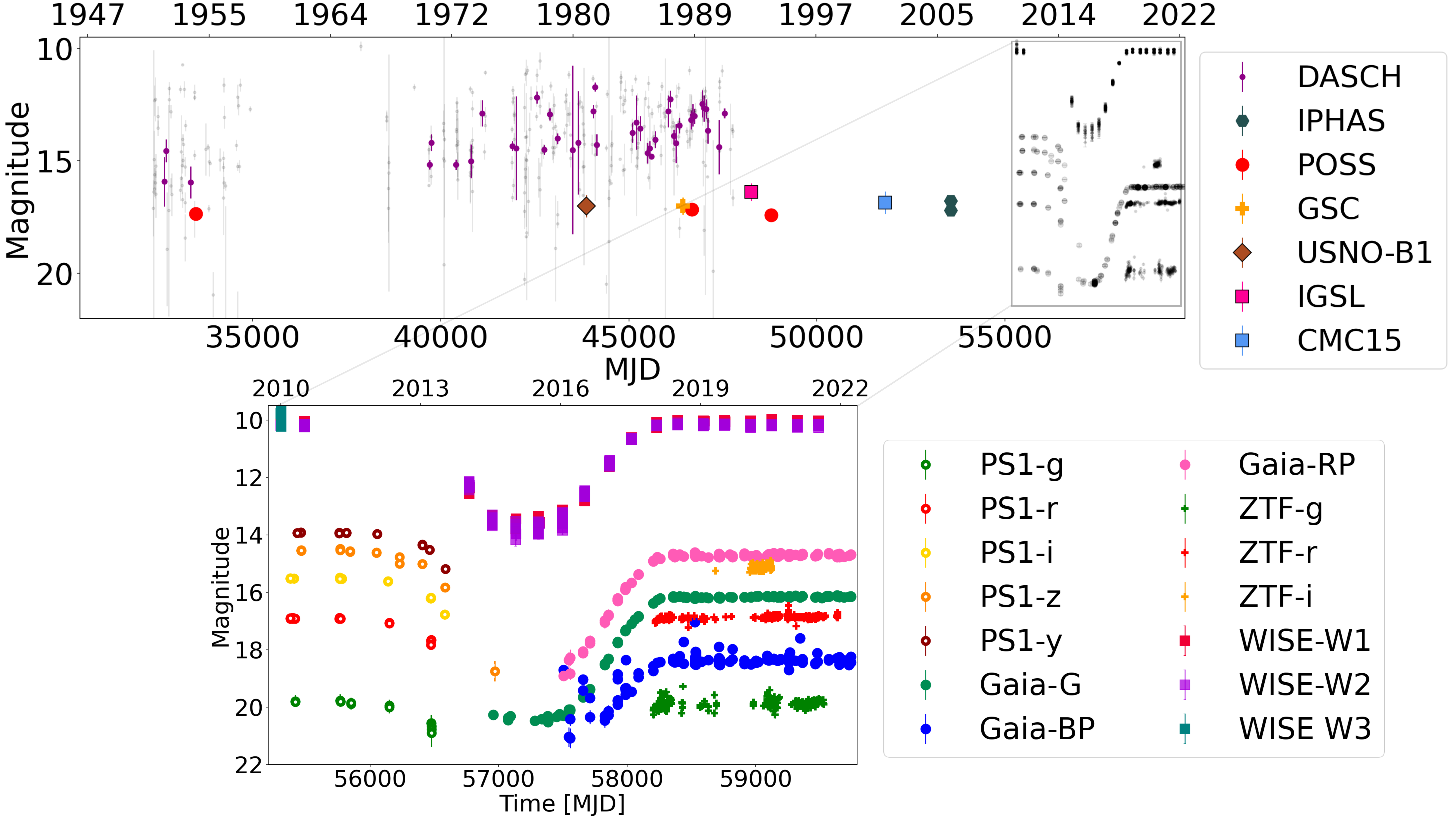}
  \centering 
  \caption{Historical optical-NIR light curve of Gaia17bpp. We correct the $PS1_{r}$ and $PS1_{g}$ to match the average magnitude baseline of the ZTF light curves.}
  \label{fig:photometry}
\end{figure*}

Outside of the primary dimming event, Gaia17bpp was also detected in the Zwicky Transient Facility (ZTF) survey \citep{2019PASP..131a8002B} in the \textit{gri} broadband filters. In short, ZTF is a 47 square-degree time-domain survey utilizing the 48-inch Schmidt Palomar telescope (P48) located at the Palomar Observatory. ZTF performs a public component survey that covers the entire northern sky with a 2-day cadence in $g+r$ ($\approx$34$\%$ of P48 time) and a 1-day Galactic plane survey in $g+r$ ($\approx$6$\%$ of P48 time). A high cadence $\approx$2500 square degrees ($3g + 3r$ per night) was conducted for collaboration time including a 4-day $i-$band survey. On average the survey covers $\sim$3750 square degrees per hour with a median limiting magnitude of r$\sim$20.6 mag, g$\sim$20.8 mag, and i$\sim$19.9 mag 5$\sigma$ limiting magnitude in 30-second exposures \citep{DekanyNew}. For in-depth coverage of the scientific goals and survey strategy, we suggest to the reader review \citet{ztf_paper1} and \citet{2019PASP..131g8001G}. No other ZTF or GPSA has been issued within 20 arcseconds of Gaia17bpp. In Figure \ref{fig:photometry}, we show the optical to IR mosaic light curve of Gaia17bpp.

\subsection{Spectroscopy}\label{sec:Spectrum}

\begin{figure*}
  \includegraphics[width=0.9\textwidth]{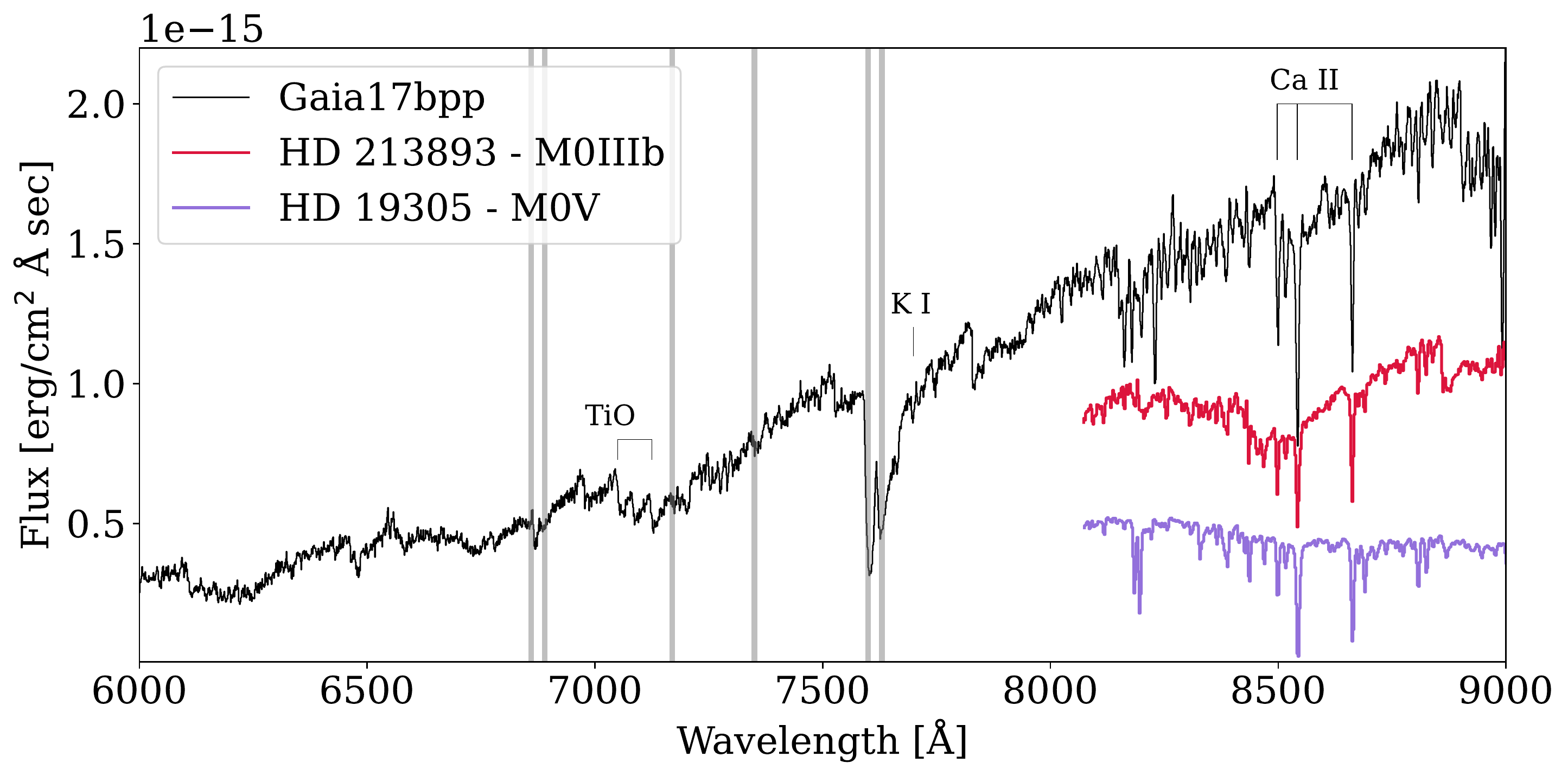}
  \caption{The optical spectrum of Gaia17bpp using the APO 3.5-meter KOSMOS spectrograph using the high-red grating. The uncorrected telluric lines have been marked with gray solid lines.  We compare the Gaia17bpp spectrum to HD 19305 (M0V) and HD 213893 (M0IIIb) standards \citep{2009ApJS..185..289R}.}
  \label{fig:spectrum}
\end{figure*}

Gaia17bpp was observed with the Apache Point Observatory (APO) ARC 3.5-meter telescope using a medium dispersion spectrograph KOSMOS \citep{2010SPIE.7735E..4LS}. We used the red high slit configuration with wavelength coverage between 6150-9800 Å. On September 20, we obtained one thirty-minute exposure with clear sky conditions. Our data reduction followed standard long-slit spectroscopy procedures and reduced the optical spectrum using the open-source Python package \texttt{PyKOSMOS} \citep{davenport_james_2021_5120786}. In Figure \ref{fig:spectrum}, we show the final reduced optical spectrum of Gaia17bpp without correcting for the telluric absorption features. In the same figure, we also include the NIR spectra of an M0 dwarf and giant, HD 19305 (M0V) and HD 213893 (M0IIIb) standards \citep{2009ApJS..185..289R}. Upon inspecting the optical spectrum of Gaia17bpp, we do not find any immediate anomalous spectroscopic features. Generally, we identified several absorption lines such as a prominent Ca II triplet at $\lambda \lambda$8498, 8542, and 8662 Å, and clear 7050Å TiO bands, including other molecular lines near $\lambda \lambda$8160Å. We did not identify other significant narrow emission lines that appear in the spectrum, except for a very weak H$_{\alpha}$ emission line. The absence of the Calcium Hydride (CaH$_{2}$) $\lambda \lambda$ 6382/6830Å, Na I doublet 8183/8195Å, and the weak K I 7665/7699Å lines indicate the presence of an evolved giant star \citep{2007ApJ...657..511A, 1986PASP...98..467S}. The identified spectroscopic features are generally consistent with an ordinary late-type cool K-M class type star. In Section \ref{sec:PrimaryStar} we discuss in depth how the interpretation of such absorption lines is consistent with the classification of an RGB star as the primary source. The collected spectrum is available for further inspection on the Transient Naming Server (TNS) under the source name AT2017exj\footnote{\href{https://www.wis-tns.org/object/2017exj}{https://www.wis-tns.org/object/2017exj}}. 

\subsection{Follow-up Photometry}
On May 7, 2023, we performed follow-up photometry on Gaia17bpp with the APO 3.5-meter Astrophysical Research Consortium Telescope Imaging Camera (ARCTIC) \citep{2016SPIE.9908E..5HH} using the SDSS-u filter. The seeing was approximately 1-arcsecond throughout the entire night. We note that the Moon was $\sim$91$\%$ full, with a large $>$80 deg separation from our target. We observed the Gaia17bpp for a total of 23, 3-minute images to eliminate sky brightness background and noise levels. We did not perform dithering. We reduced our data following standard calibration procedures. We assumed a photometric-zero-point offset for the SDSS-u filter of 24 mag. We used the Python open-source tool, \texttt{photutils}, to perform the standard reduction of the SDSS-u band science images including our calibration files. No source was detected within 10 arcseconds of the Gaia17bpp sky position. Based on the 5$\sigma$ identified sources we found in our deep ARCTIC images, the distribution was roughly Gaussian with an average SDSS-u band magnitude of $\sim$22 mag.\\ 

\section{Primary Star}\label{sec:PrimaryStar}
\subsection{Spectral Energy Distribution}

We attempt to infer the physical parameters of the primary star. Given the rich multi-wavelength archival photometry, we perform a stellar SED modeling to constrain the properties of the primary star using open-source SED fitter \texttt{Ariadne} \citep{astro_adri}. \texttt{Ariadne} deploys a suite of stellar atmosphere model grids by performing a convolution of synthetic SED atmospheres and broadband photometry to constrain the effective temperature, metallicity, distance, radius, and extinction of a given SED. It performs a dynamic Bayesian multi-nested sampling algorithm through \texttt{dynesty} \citep{2020MNRAS.493.3132S}, and amalgamates each stellar atmospheric model using a Bayesian Model Averaging (BMA) to finally estimate the best-fitted model parameters of a given primary star through a weighted average. In our case, we began by including all the observed broadband photometry outside the dimming event. In Table \ref{table:pre-eclipse-photometry} we summarize all the observed magnitudes outside the primary dimming event from all epochs before 2013 or after 2018. For the fitting, we assumed a \cite{1999PASP..111...63F} extinction law with A$_{V}$=5.3 assuming a Galactic R$_V$=3.1 value. In Table \ref{table:prior-posterionr} we show the priors and posterior distributions used for this analysis.

\begin{figure*}
  \includegraphics[width=0.9\textwidth]{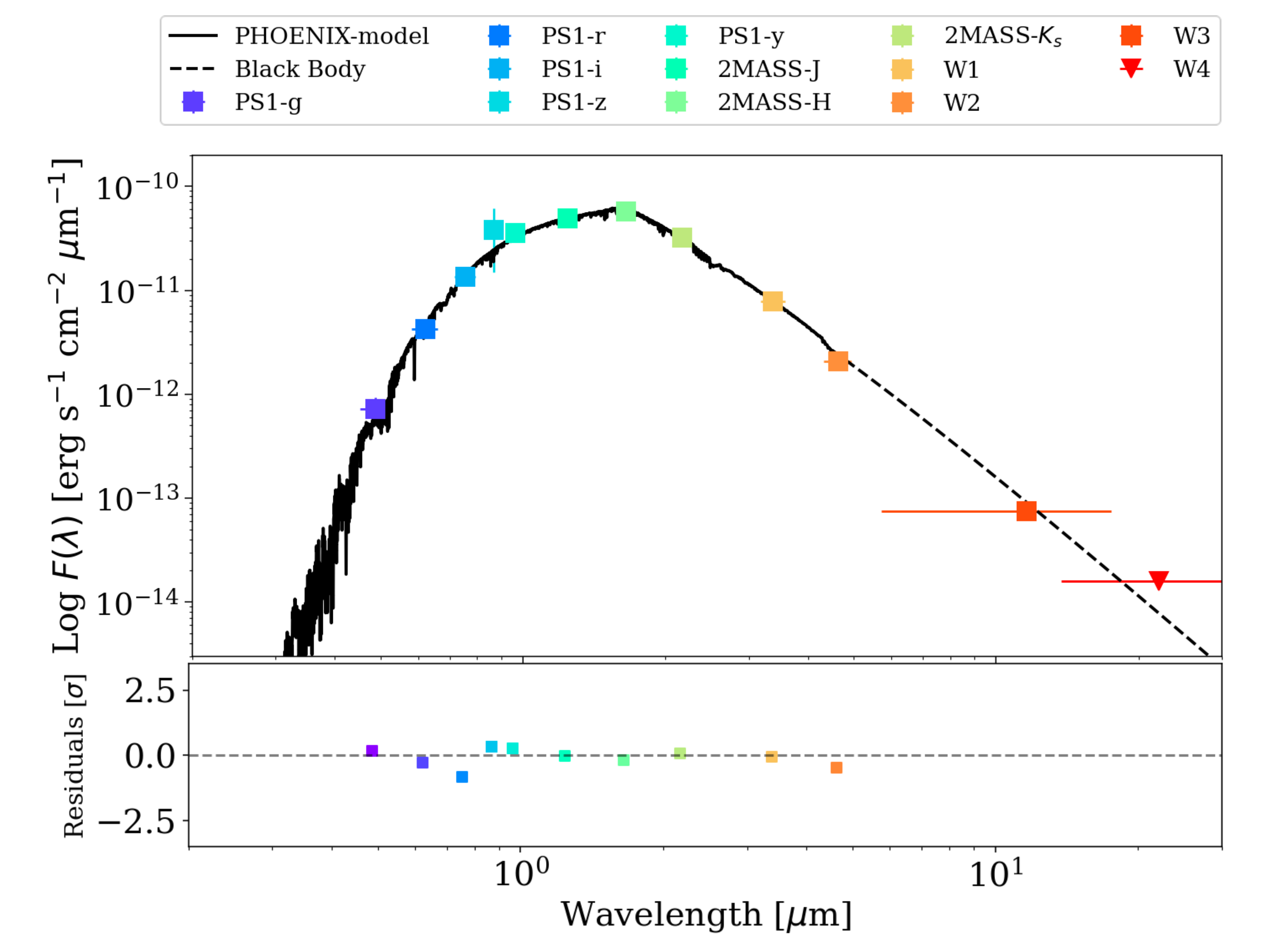}
  \centering 
  \caption{Spectral energy distribution fit (black synthetic PHOENIX spectrum) to the obtained broadband photometry outside the Gaia17bpp dimming event. We also include a simple black body fit to the photometry accounting for the effects of extinction at the line of sight. The bottom panel represents the residual levels from the PHOENIX model. We note that the SED fitting model does not extend to the W$_{3}$ detection and W$_{4}$ upper limit.}
  \label{fig:sed_fit}
\end{figure*}

\begin{figure*}
\centering
  \includegraphics[width=0.7\textwidth]{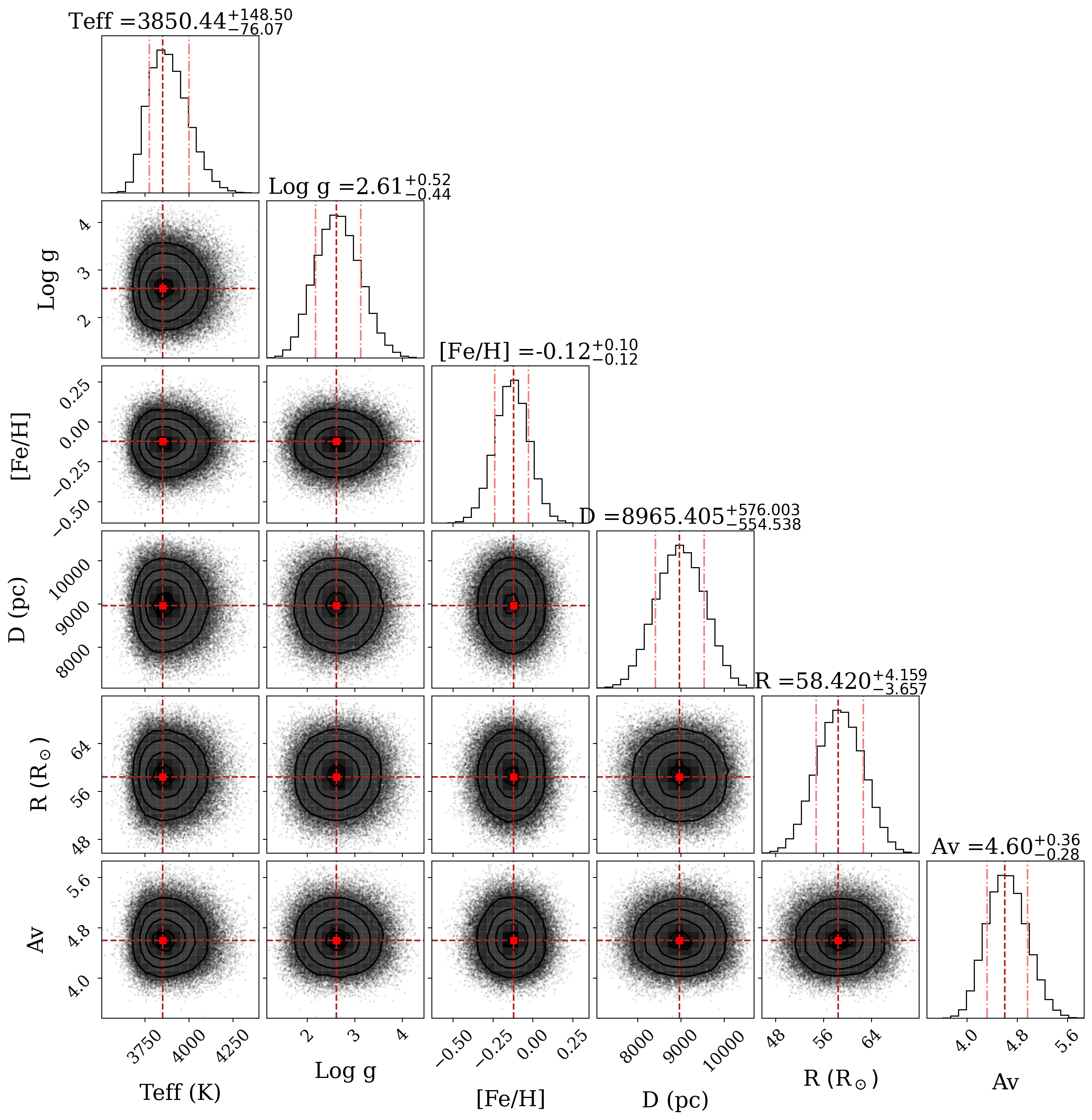}
  \caption{Corner plot of the posterior distribution from the nested Monte Carlo sampling using \texttt{Ariadne}. The dashed lines are the median and $2\sigma$ confidence interval of each distribution.}
  \label{fig:SED_corner}
\end{figure*}

\begin{figure}
  \includegraphics[width=0.45\textwidth]{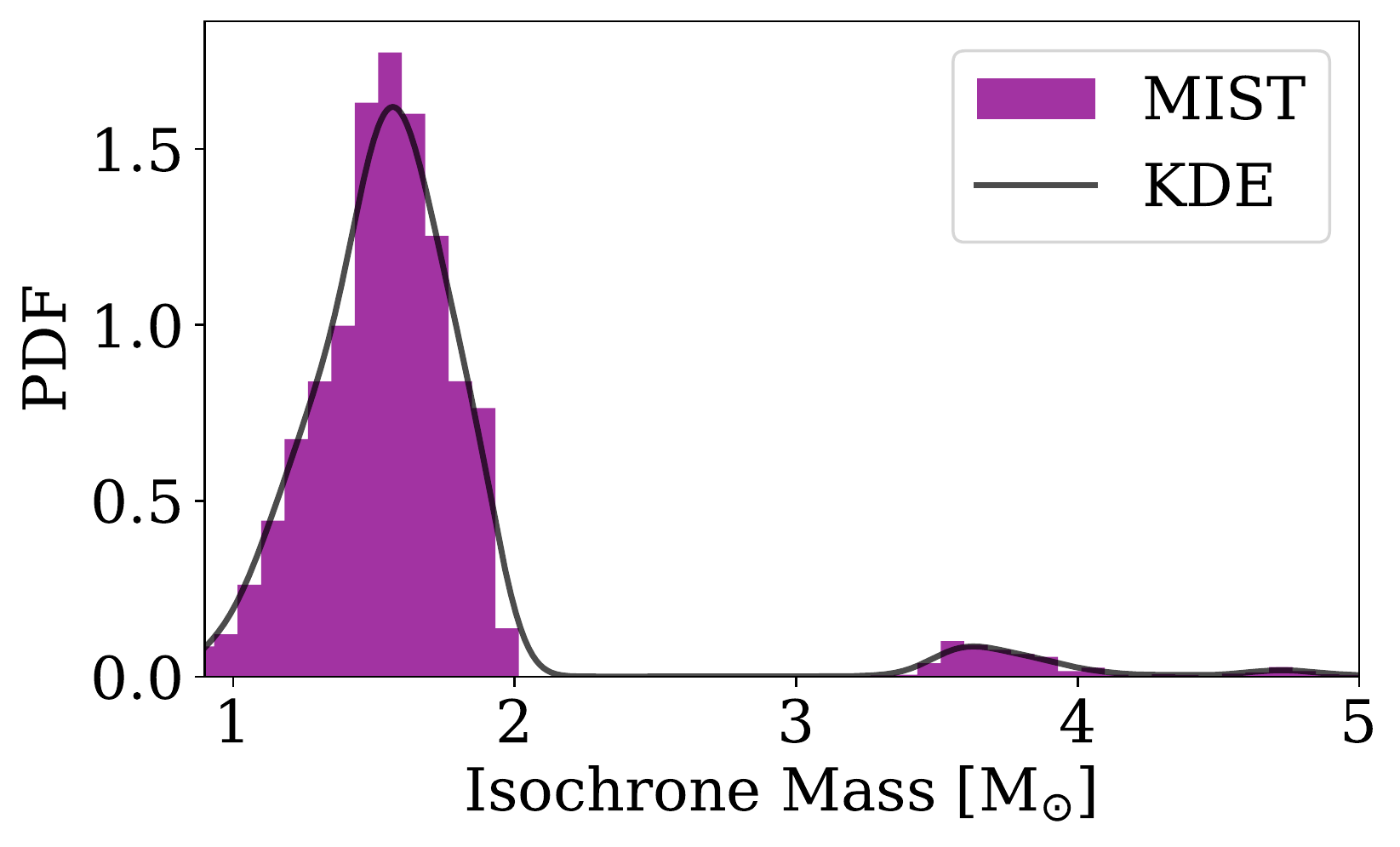}
  \caption{MIST isochrone mass posterior distribution histogram including a kernel density estimation (black line). }
  \label{fig:mass_kde}
\end{figure}

In Figure \ref{fig:sed_fit} we show the synthetic BMA PHOENIX atmospheric model that best fits our broadband photometry up to 4.6 $\mu$m. In our SED plot, we also include the W$_{3}$ detection and W$_{4}$ upper limit of the primary star, however, the SED fitter does not fit those detections. We extended the synthetic PHOENIX model continuum by fitting a black body function at the best-fitted temperature, radius, and distance  (shown in dashed lines in Figure \ref{fig:sed_fit}). The average fitted model shows convergence on well-constrained posterior distribution seen in Figure \ref{fig:SED_corner} including their 25th, 50th, and 75th percentiles. The posterior distance distribution is in good agreement with the photo-geometric distance estimation. According to the physical estimates we derive from the SED fitting, we find the primary star to be consistent between K5-III or M0-III giant star according to the classification regime considering its effective temperature, radius, and surface gravity \citep{2017A&A...605A.111C, 1998NewA....3..137D, 1996AJ....111.1705D}. The interpretation of a cool giant star is also consistent with the identified spectroscopic features discussed in Section \ref{sec:Spectrum}. Using the maximum likelihood from the BMA, \texttt{Ariadne} also fits in the background MIST isochrones given the derived luminosity and temperature to estimate the mass. In Figure \ref{fig:mass_kde} we show both the posterior distribution of the resulting stellar mass, including the kernel density estimation. It is evident that the isochrone mass posterior distribution is bimodal with a higher probability density for a lower mass stellar model with a median of 1.56 M$_{\odot}$. \\

\begin{table}[ht]
\centering 
\begin{tabular}{c c  c  c} 
\hline\hline 
Survey/Band & Magnitude & Magnitude Error\\ [0.05ex]
\hline 
2MASS-H & 10.7420 & 0.0210 \\
2MASS-J & 12.0080 & 0.0220   \\
2MASS-Ks & 10.3080 & 0.0180 \\
\hline
PS1-g    &  	 19.5110 	&  0.0332   \\
PS1-i    &  	 15.3795 	&  0.0130   \\
PS1-r    &  	 17.0549 	&  0.0089   \\
PS1-y    &  	 13.7903 	&  0.0452   \\
PS1-z    &  	 13.9440 	&  0.1893   \\
\hline
WISE-W1   &	 10.0630  &	  0.0240 \\ 
WISE-W2   &	 10.1690  &	  0.0220 \\
WISE-W3   & 9.9180    & 0.1469 \\
WISE-W4   &  8.774\footnote{2$\sigma$ upper limit.}    & 0.1469 \\
\hline \hline 
\end{tabular}
\caption{Compiled broadband photometry of Gaia17bpp outside the dimming event. We assume that the intrinsic stellar SED be the same before or after the dimming event. \label{table:pre-eclipse-photometry}} 
\end{table}

\begin{table}[ht]
\centering 
\begin{tabular}{c c c} 
\hline\hline 
  Parameter & Prior & Posterior \\ [0.05ex]
\hline 
T$_{\text{eff}}$ & $\mathcal{U}$(3000, 5000) & 3927$^{+155}_{-108}$ K \\
log(g) & $\mathcal{U}$(1, 5) & 2.7$^{+0.6}_{-0.5}$ cm sec$^{-2}$ \\
$\text{[Fe/H]}$ & $\mathcal{U}$(-1, 0.5) & -0.1$^{+0.1}_{-0.1}$ dex \\
R$_{\text{ad}}$ & $\mathcal{U}$(5, 100) & 59.7$^{+3}_{-5}$ R$_{\odot}$ \\
D$_{\text{ist}}$ & $\mathcal{U}$(7000, 10$^4$) & 9250$^{+394}_{-770}$ pc \\
A$_{V}$ & $\mathcal{U}$(2, 5) & 4.75$^{+0.4}_{-0.3}$ mag \\
\hline \hline 
\end{tabular}
\caption{The table contains the prior and posterior distributions from the primary star SED analysis. For the posterior distribution, we report the median and the 2$\sigma$ posterior distribution range.\label{table:prior-posterionr}} 
\end{table}

\hfill \break

\subsection{Primary Star Variability}
Since the primary large amplitude dimming event, the Gaia17bpp system thus far has not exhibited any other large-scale variability beyond an overall 0.05 mag baseline scatter that was calculated using the ZTF-g, ZTF-r, and Gaia $G$ light curve at all observations after 2018. We searched the ZTF-g, ZTF-r, and Gaia $G$ light curves for periodic or quasi-periodic variability outside the dimming event. We ran a single-band Lomb-Scargle Periodogram (LSP) \citep{Lomb, Scargle} using \texttt{gatspy} \citep{2015ApJ...812...18V} using a two-component Fourier model. Across each photometric bandpass, we got inconsistent periods at maximum power which is likely due to the difference in cadence between each filter and survey. Despite the identification of periods exceeding 90 days in our LSP analysis, indicative of pulsating red giants in accordance with Kepler light curves \citep{2017A&A...605A.111C}. We calculated the significance of the peak through a bootstrap extrapolation and found that it fell below the 2$\sigma$ threshold.



\section{Dimming Event Properties}\label{sec:DimmingEventProperties}

\subsection{Light Curve Properties: Optical to IR}

Given the remarkably long duration of the dimming event of Gaia17bpp, there are five photometric bandpasses that extensively covered the second half of the dimming event. The bottom panel in Figure \ref{fig:photometry} shows how stitching together detections from multiple surveys and photometric bandpasses reveals almost the entirety of the dimming event except for the ingress. Both WISE and NEO-WISE light curves did not show any evidence of large amplitude variability during both the occultation and dimming phases. Based on a 30-day median sliding window, we found that the light curve did not vary more than 0.2 magnitudes across both the W$_1$ and W$_2$ magnitudes. The W$_1$ and W$_2$ light curves returned back to their median baseline brightness after the dimming event. Similarly, the Gaia $G$ light curve did not exhibit any structure during the dimming event. Due to the sparse coverage of all presented photometric measurements, we recognize that there might exist variability on shorter timescales that are not captured. 

\begin{figure*}
  \includegraphics[width=0.7\textwidth]{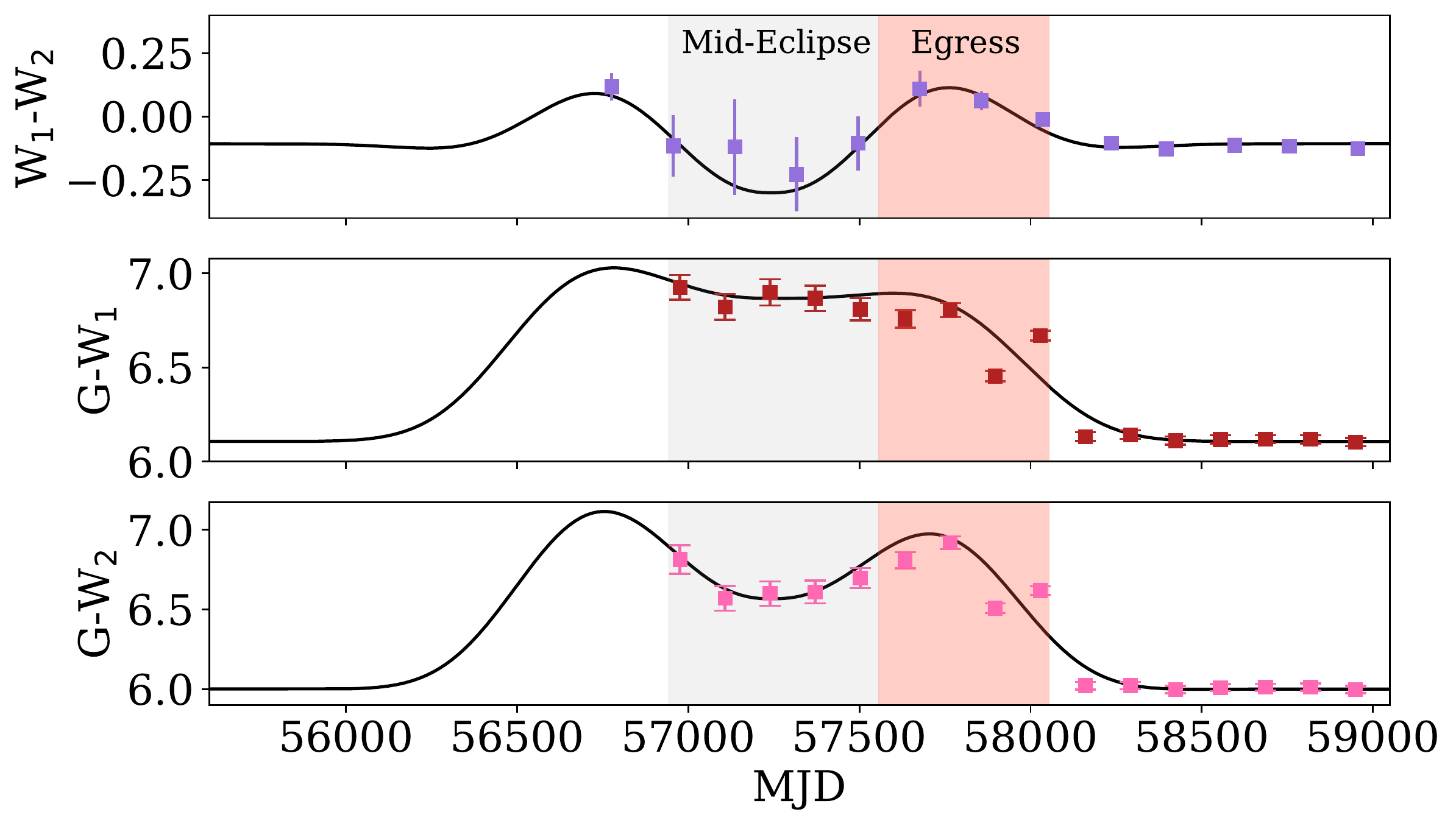}
  \centering 
  \caption{Color evolution of Gaia17bpp in the infrared and infrared-optical photometry. We highlight in gray the primary region of the dimming event and in red the egress of the dimming event. In each panel, the underlying solid black line model was produced from a high-order Gaussian function described in Section \ref{sec:DimmingEventProperties}.}
  \label{fig:color_evolution}
\end{figure*}

Perhaps the more puzzling property of the Gaia17bpp system is the variation in colors across different epochs. In Figure \ref{fig:color_evolution} we show the color evolution of the dimming event including the non-dimming phase in three photometric bandpasses. In order to get the same epochs for the WISE and Gaia photometry, we binned the detection by a running median of 100 days. We first begin to note the constant color during the quiescent phase found in any epoch beyond January 2019, this constant color profile is the inherent color of the primary star including Galactic foreground extinction. We do not observe any significant color variations within this phase. On the other hand, both the egress and bottom of the dimming event show substantial color variation. For example, in the top panel of Figure \ref{fig:color_evolution} we notice a blue 0.25 mag bump (henceforth blue-bump) in the color profile at the bottom of the dimming event. This is also accompanied by a slight reddening during the ingress and egress detections. The Gaia-WISE colors also suggest a similar evolution and a small blue bump at the minimum of the dimming event at different rates. The last two detections are also somewhat curious with a 200-day color 0.1 mag variation from blue to red, however, it is difficult to tell if these are spurious detections or related to the dimming mechanism. 

Given the five unique photometric filters, we have that captured the egress of the dimming phase, we suspect that the shape of the light curves can possibly change from optical to infrared assuming that the obscuring process does not have an underlying grey spectrum. To test this hypothesis, we fit the overall properties of the light curve with a high-order Gaussian function:
\begin{equation}
    y(t | A, \mu, \sigma, \gamma) = A \exp \left(-\left(\frac{\left(t-\mu\right)^2}{2 \sigma^2}\right)^\gamma\right)
\end{equation}
with amplitude (A), mean ($\mu$), standard deviation ($\sigma$), and overall order of the Gaussian function and Gaussian wing profile ($\gamma$) that controls the flatness of the beak peak wings of the distribution. The power of the underlying high-order Gaussian is the ability to control the flatness of the Gaussian peak since we can visually tell that the bottom of the dimming event was flat. To account for any unwanted correlated noise measurements, we simultaneously model the underlying correlated noise with a Gaussian Process with \texttt{George}\footnote{\textcolor{blue}{\href{https://george.readthedocs.io/}{https://george.readthedocs.io/}}} \citep{2015ITPAM..38..252A} including the mean model parameters via Markov chain Monte Carlo (MCMC). We use \texttt{Emcee} \citep{2013PASP..125..306F} that implements the Metropolis Hasting algorithm used to constrain the posterior distribution. The correlated noise that might be present in the multi-band light curves was captured by a Matérn-3/2 covariance function. We simultaneously fit the GP hyperparameters, including the underlying model. 

\begin{figure}
  \includegraphics[width=0.5\textwidth]{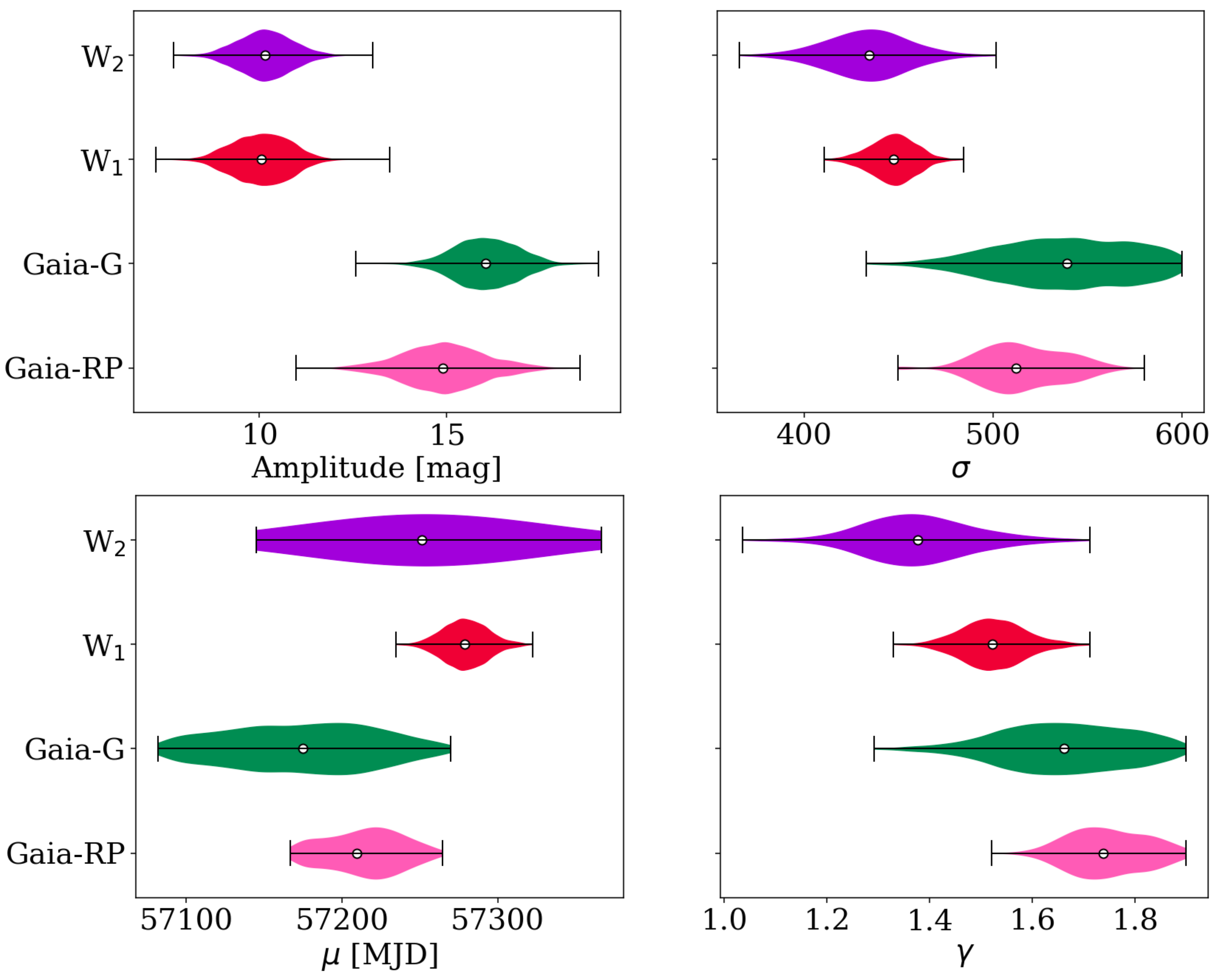}
  \centering 
  \caption{Violin plot showing the posterior distribution of the high-order Gaussian function we fitted for each photometric bandpass. The white dot for each band represents the median of each posterior distribution. }
  \label{fig:hyperpars}
\end{figure}

Overall, the high-order Gaussian model provided a good fit for the RP, Gaia $G$, W$_{1}$, and W$_{2}$ light curves. Based on our GP analysis, we found clear deviations from the overall shape of the light curves across each bandpass. For example, in Figure \ref{fig:hyperpars} we show the posterior distributions for each photometric filter, for the model mean time and wing shape of the Gaussian profile. It is noticeable how the peak of the $\gamma$ posterior distribution shifts from the optical 1.6 to 1.3 in the infrared, suggesting that the infrared light curves are less flat compared to the optical ones. Thus it is possible that the light curve profile of the Gaia17bpp dimming event is modestly asymmetric. Based on the wide distributions we obtained from the model mean epoch, all light curves had a minimum flux of around June 2016 (57200 MJD) within a margin of error of $\sim$400 days. We could not successfully place firm limits on the BP light curve since it suffers from a large scatter. 

Given the poor photometric coverage at the minimum of the dimming event, we do not have enough multi-band photometry to perform any SED inference. We attempted to fit a simple black body model to the BP, G, RP, PS1-y, W$_{1}$, and W$_{2}$ detections and found that the temperature was consistent with the findings of the dimming phase at 3800 K. We only notice that the dimming detections are roughly equivalent to the non-dimming detections scaled down by some factor. 

\section{Discussion} \label{sec:Discussion} 

\subsection{Intrinsic Variability Scenarios} \label{sec:int_var}
Month- to year-long dimming events have been identified from rare hydrogen-deficient carbon-rich stars, R Coronae Borealis (RcB) type variables that eject large carbon-dust clouds. RcB stars are known for their erratic and asymmetric dimming profiles and are dominated by mostly helium, nitrogen, and carbon enrichment in their atmospheres \citep{2002AJ....123.3387D}. We found no He-I emission or P-Cygni profile in our optical spectrum. RcB stars are also known to exhibit mid-IR excess due to warm circumstellar dust shells found around them \citep{2012A&A...539A..51T}. No carbon absorption features or other spectroscopic RcB signatures were found in our optical spectrum. Considering the stellar SED of Gaia17bpp, we also do not find any evidence of mid-IR excess. We examined the WISE color-color magnitude diagram of the primary star to see if the colors were consistent with known RcB stars. We compared the (W$_{1}$-W$_{2}$)$_{0}$, (W3-W4)$_{0}$ with values -0.17 and 1.1 respectively, and found that the observed color is 0.5 mag away from where typical RcB stars have been found in the WISE catalog \citep{2020A&A...635A..14T}. Given the additional absence of any other erratic variability and the non-asymmetric light curve of Gaia17bpp, we find it unlikely that it is an RcB candidate. 

Red supergiants (RSG) are also known to exhibit large amplitude dimming events that can last up to a few months to a year \citep{2022ApJ...936...18D, 2007ApJ...660..301M}. One particular case study is the great dimming event of Betelgeuse which exhibited a deep 2-month-long dimming event. Initial interpretations noticed the temperature of Betelgeuse had cooled down to 3600 K, and suggested surface convection effects caused the dimming event. \citet{2020ApJ...891L..37L} discussed that a large and cool convection cell on Betelgeuse would have caused strong TiO bands with a substantially lower temperature, instead, the leading hypothesis is an episodic mass loss event with large grain circumstellar dust. Our current analysis does not favor an RSG classification for Gaia17bpp given its modest radius and low-mass estimates based on the SED. Similarly, the duration of the dimming event we observe is not compatible with the timescales currently reported in the literature on dimming RSG stars \citep{2006MNRAS.372.1721K}. We find it unlikely that the Gaia17bpp dimming mechanism was caused by the ejection of a dust cloud around RSG stars.

\subsection{Extrinsic Variability Scenarios}\label{sec:extVar}

Our current available data do not provide evidence that Gaia17bpp is a compact X-ray binary or an energetic cataclysmic variable (CV). We searched through known X-ray catalogs (ROSAT 2RXS, XMMDR10, XMMSL2, Chandra CSC2, swift2SXPS) and found no sources at the location of Gaia17bpp except for a 2$\sigma$ upper-limit from Swift/XRT count rate at 0.36 counts per second. We derived an X-ray limiting luminosity of approximately 10$^{35}$ erg sec$^{-1}$ at the distance of the source\footnote{To calculate the limiting X-ray luminosity we utilized NASA HEASARC WebPIMMS calculator. For our calculation, we assumed an H I column density at the line of sight to be 6.9 cm$^{-2}$ assuming a median distance of 8462 parsecs. We assumed no intrinsic HI column density.}. Given the relatively shallow upper X-ray flux limit at this line of sight, it is possible that Gaia17bpp is an active X-ray source but has been missed due to the lack of deeper limits. Other common telltale signs of accretion are usually seen in H$\alpha$ emission from an ionized disk \citep{1980ApJ...235..939W}. As seen in Figure \ref{fig:spectrum} we do not see any narrow H$\alpha$ or other emission lines of ongoing accretion. Given the lack of erratic photometric variability, no X-ray detections, and no evidence of narrow emission lines in the optical spectrum at the time, we do not believe the dimming of Gaia17bpp results from the activity in an accretion disk.

Young stellar objects (YSO) are systematically known to display deep and complex dimming events on timescales of months to years called dipper events \citep{2018AJ....156...71C}. In most cases, YSO disks are also luminous X-ray and radio sources due to the strong magnetic fields and high temperatures from ongoing accretion \citep{1999ARA&A..37..363F}. Our archival data does not support a YSO scenario given the lack of infrared excess, the lack of bright nearby radio or X-ray sources. Since the typical timescales of YSO disks are a few Myr, we do not expect to find such a young companion near an evolved giant star. Finally, the sky position and estimated distance of Gaia17bpp is not near any known star-forming regions.

\subsection{Occulting Disk Model}\label{sec:disk_model}
As discussed in Sections \ref{sec:int_var} and \ref{sec:extVar}, there is no currently known intrinsic or extrinsic dimming mechanism that can cause giant stars to smoothly dim for prolonged periods. We now turn our attention to the most likely cause of the dimming event which would be driven by the occultation of a disk that explains the dimming event of Gaia17bpp.

The smooth and asymmetric dimming event of Gaia17bpp can plausibly be explained by the eclipsing of an oblate occulting disk similar to that of $\epsilon$-Aur analog systems. For example, \citet{2019MNRAS.485.2681R} showed that EPIC 204376071, a young M-type star, was obscured by an oblate tilted dusty disk. To compute the transiting models, they used a modified version of \texttt{pyPplusS} \citep{2019MNRAS.490.1111R} that computes light curves for oblate spherical exoplanets that possess rings and can account for both uniform and limb-darkened scenarios which are performed using the Polygon-Segments algorithm. \cite{2019MNRAS.485.2681R} expanded this tool by adding a full opacity layer to the rings to simulate the transit of an oblate disk. 

\begin{table}[ht]
\centering 
\begin{tabular}{c c c} 
\hline\hline 
  Parameter & Prior & Posterior \\ [0.05ex]
\hline 
R$_{disk}$ $[R_*]$ & $\mathcal{U}$(0.1, 10) & 6.5$^{+4.0}_{-2.8}$\\
b & $\mathcal{U}$(0, 10) & 2.6$^{+4.9}_{-2.6}$\\
i $[rad]$ & $\mathcal{U}$(0, $\pi$) & 1.0$^{+0.3}_{-0.7}$\\
$\phi$ $[rad]$ & $\mathcal{U}$(0, $\pi$) & 3.1$^{+0.1}_{-0.2}$\\
v $[km \ sec^{-1}]$ & $\mathcal{U}$(10$^{-4}$, 10) & 0.005$^{+0.002}_{-0.001}$ \\
dx & $\mathcal{U}$(0, 200) & 114.0$^{+79.5}_{-63.8}$\\
\hline \hline 
\end{tabular}
\caption{The table contains the prior and posterior distributions from the oblate disk modeling. For the posterior distribution, we report the median and the 1$\sigma$ range.\label{table:params-disk}} 
\end{table}

\begin{figure}
  \includegraphics[width=0.5\textwidth]{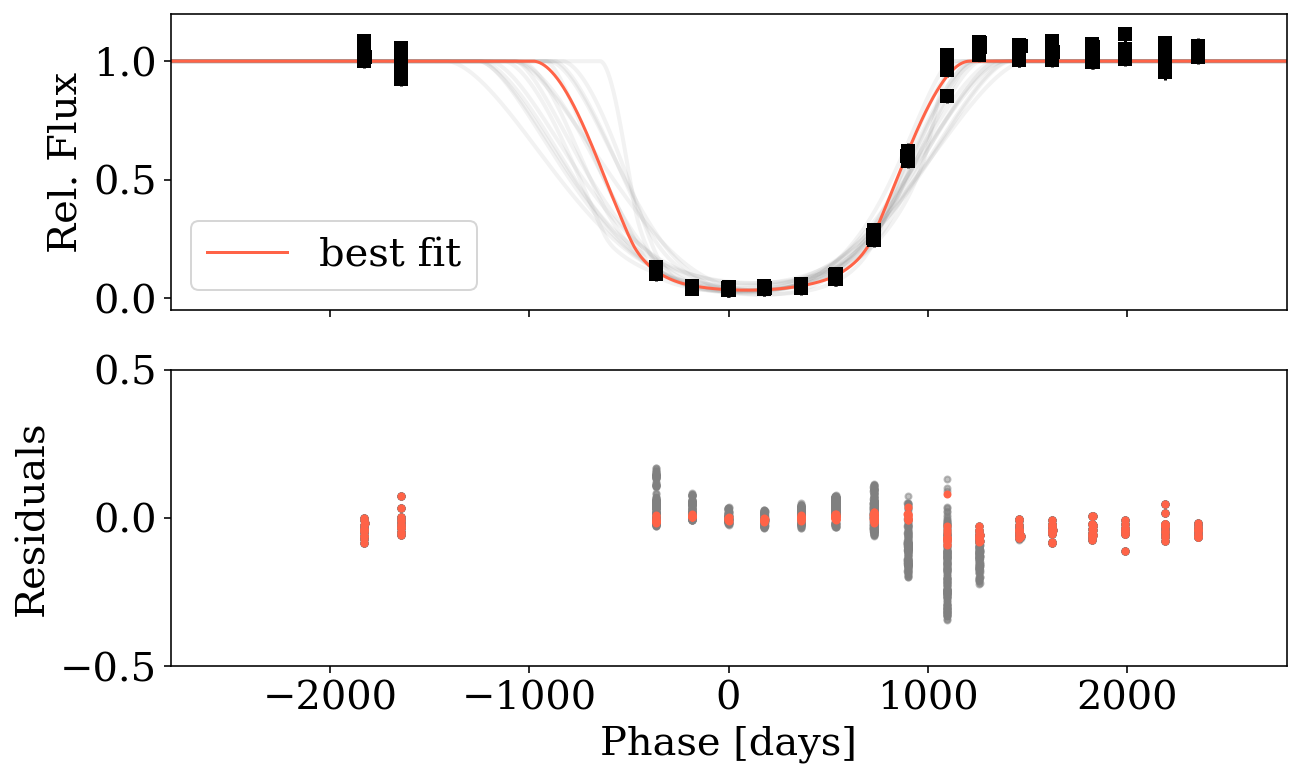}
  \centering 
  \caption{Gaia17bpp WISE-1 light curve fitted by an oblate solid disk model (top) and model residuals (bottom). In pink we show the maximum likelihood fit to the data. The gray points represent random draws from the posterior distribution.}
  \label{fig:disk_fit_model}
\end{figure}

We attempted to apply this suite of models to the W$_{1}$ light curve of Gaia17bpp since it contains detections from all phases of the event. One of the key modifications we made to the existing code was to also include in the fitted light curve the limb-darkening coefficient. Given the priors we obtained from our SED analysis, we only consider the c$_2$ coefficient of the quadratic limb-darkening equation since the first linear term is very small\footnote{To calculate a preliminary limb-darkening coefficient we used the Exoplanet Characterization Toolkit calculator (\href{https://exoctk.stsci.edu/limb_darkening}{https://exoctk.stsci.edu/limb\_darkening}).}. For the purpose of this study, we only fit a model of an oblate solid disk model with seven parameters: disk radius in radii of the primary star (R$_{\text{disk}}$), impact parameter (b), orbital inclination (i), the tilt angle of the disk ($\phi$), transverse speed of the disk across the host star (v), time shift (dx), and finally the quadratic polynomial limb-darkening term (c$_2$). We explored a variety of different model parameters that could produce a 7-year-long dimming event. We found that the models that could produce such long events needed to have low transverse velocities that likely indicate the large semimajor axis between the occulter and primary star. We proceeded to perform an MCMC implementation using a standard Gaussian likelihood function. In the limitation of poor constraints on the geometry of the system, we assumed broad uniform priors for all parameters. In Figure \ref{fig:disk_fit_model} we show the model fit against the data. While the disk light curve model performs overall a reasonably fit to the data, we do find some weakly correlated noise within the residuals (bottom panel of Figure \ref{fig:disk_fit_model}). Considering the limited constraints at our disposal, and the prior orbital configuration of this system, it remains very challenging to constrain based on the current data. Nonetheless, we found that the semi-major axis of the disk is roughly 1.4 AU, including the angle between the semi-major axis and the direction of motion across the sky to be inclined by $\sim$178 degrees. It is thus possible that we are seeing the disk obscuring the star edge-on since we also find the impact parameter to be small (b$\sim$0). The marginalized posterior distribution also suggests that the transverse velocity is low, approximately at 0.005 km sec$^{-1}$. In Table \ref{table:params-disk}, we summarize the marginalized posterior distribution from our analysis. We attempted to repeat the same model with the exception of introducing an additional ring to the oblate occulter with some tunable opacity term. We found that such a two-layer oblate occulter was not capable of fitting the overall shape of the light curve due to the steep transition between a solid disk and another additional optically thin layer. Based on the performance of our single oblate transiting model, we have reasonable evidence to suggest that the occulted object of Gaia17bpp must have been an extended object such as an oblate disk. However, in order to establish more robust constraints on the geometry of the disk, additional multi-band light curve modeling will be needed.

\begin{figure}
  \includegraphics[width=0.5\textwidth]{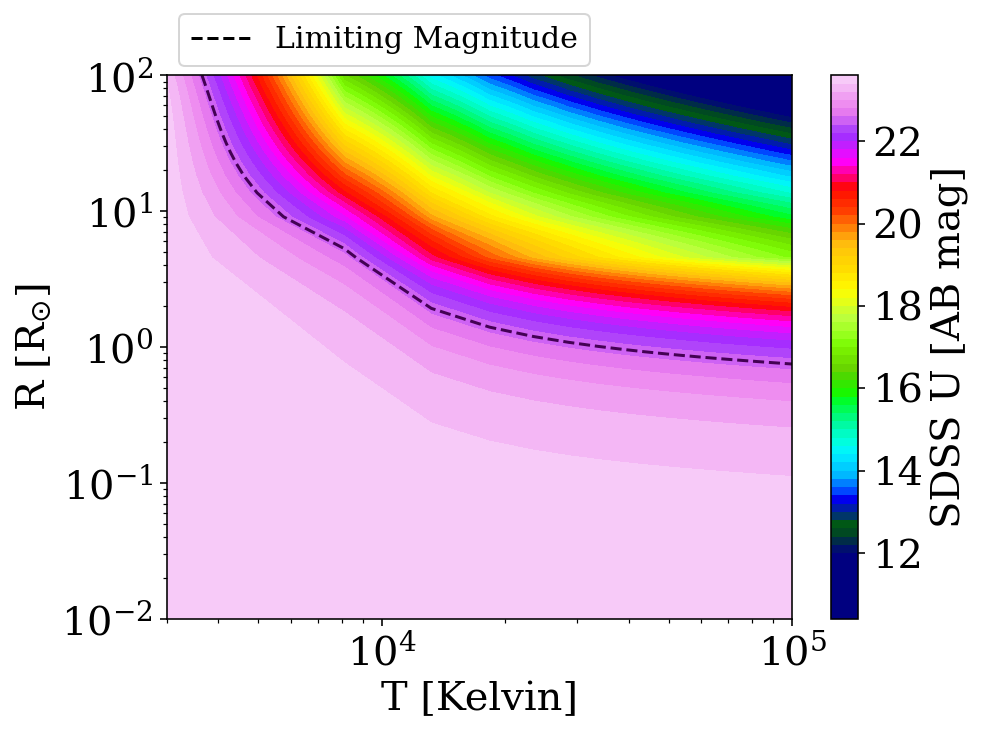}
  \centering 
  \caption{Grid of blackbody mixture model including the radius and effective temperatures of the secondary star color coded by the integrated SDSS-u AB magnitude. The dashed line represents our limiting SDSS-u band magnitude. We assume the primary star has a fixed radius of 58$R_{\odot}$, an effective temperature of 3850 K, and a median distance of 8965 pc. }
  \label{fig:blackbodyscenario}
\end{figure}

We have reasonable data to challenge the unlikely scenario of a circumbinary precessing disk such as the case of KH 15D-like analogs \citep{2022ApJ...933L..21Z, 2014AJ....147....9W}. For example, almost all known KH 15D-like analogs have been found to have characteristic shallower infrared light curves, including clear excess in infrared. Additionally, given the collected photometry, we were unable to find any short-period signals indicating the presence of a companion. One similarity between Gaia17bpp and KH 15D-like analogs is the excess of blue flux from the infrared light curves. Based on the archival data alone we are unable to identify the mechanism behind the blue-bump observed during the minimum of the Gaia17bpp dimming event. One possible scenario is the presence of another hot source contributing to the observed excess in blue flux. A second scenario is the effect of Mie forward-scattering due to the dust grain size that is comparable to the wavelength \citep{2008ApJ...681.1377S}. It is thus indicated that the characteristic grain size of the Gaia17bpp occulting disk is likely larger than the average foreground interstellar medium \citep{2016AJ....151...90A}. Compared to other possible candidates of the same nature, for example, $\epsilon$-Aur example (\cite{1991ApJ...367..278C} see Figure 5 therein) that has clear amplitude variations during the maximum eclipse, Gaia17bpp shows overall smaller variability which might shed light on the geometry and optical depth of the occulter. Gaia17bpp and $\epsilon$-Aur also show similarities in asymmetries in their light curve properties throughout different bandpasses, with $\epsilon$-Aur having an ingress dimming and somewhat steeper egress brightening phase \citep{2012JAVSO..40..618S}. It has been found that debris disks may lack infrared emission due to the non-radial distribution of mass, compared to typical protoplanetary disks \citep{2018ARA&A..56..541H}. \cite{2011AJ....142..174S} attempted to find evidence for an infrared excess or silicate features in $\epsilon$-Aur, however, the SED was found to be smooth and lacked classic dust features. It is suggested that large particles in a debris disk could account for the observed SED. Both cases of $\epsilon$-Aur and TYC 2505-672-1 have evidence to suggest the scenario of a debris disk around the secondary star. Thus this would be also in line with a lack of infrared excess that we notice with the Gaia17bpp system. Future IR studies should also investigate if Gaia17bpp contains any signatures of IR excess.

\begin{figure*}
\centering
  \includegraphics[width=0.8\textwidth]{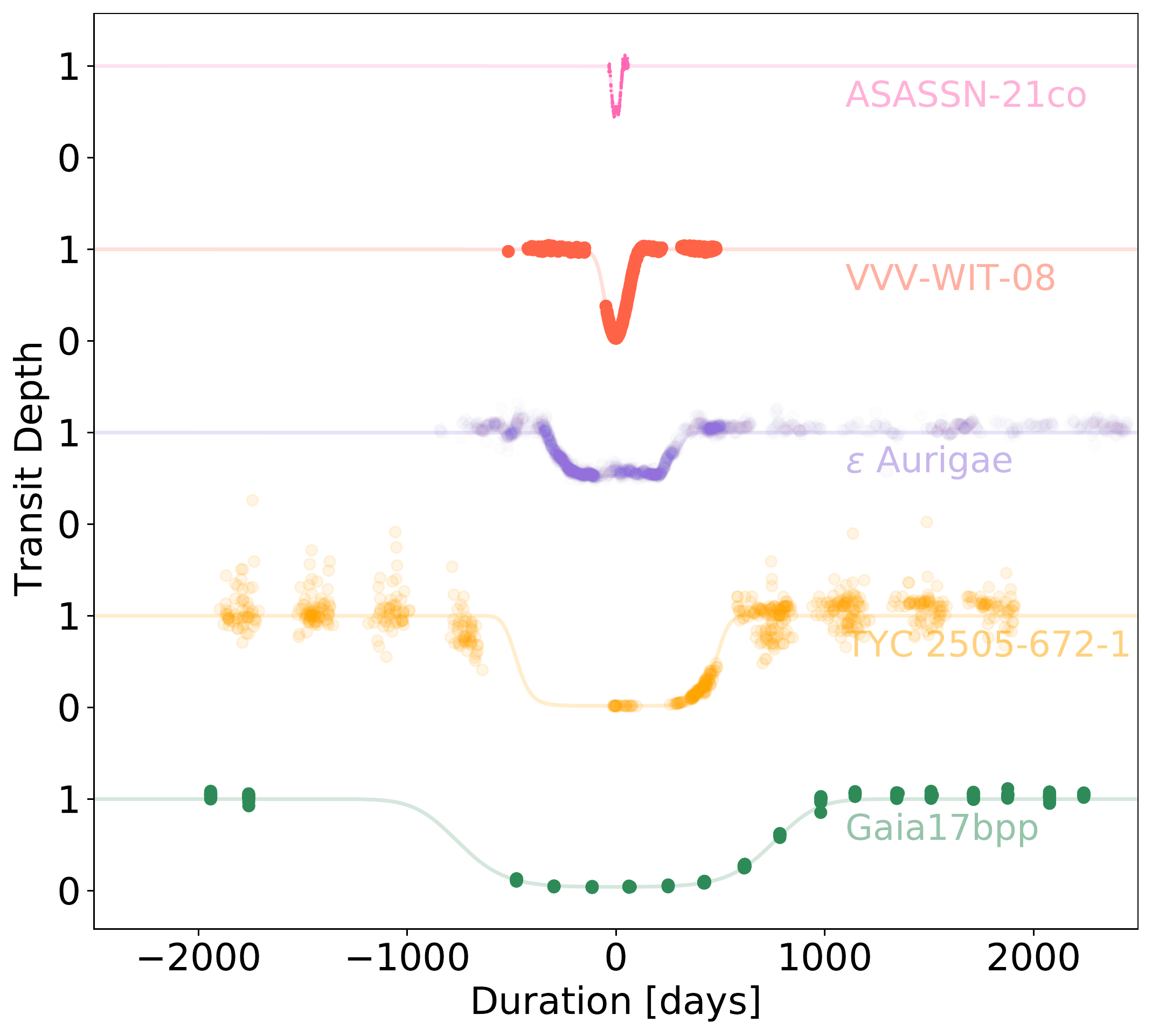}
  \caption{Normalized transit light curves of $\epsilon$-Aur-analog systems ordered by event duration. Gaia17bpp (W$_{1}$bandpass) exhibits the longest and deepest duration event. We sort each dimming event from the shortest (top) to the longest (bottom).}
  \label{fig:comp_dips}
\end{figure*}

Given Gaia's superb micro-arcsecond resolution, if a luminous companion existed it would likely be detected by Gaia. For instance, assuming a 1.5 M$_{\odot}$ and a minimum orbital period of 66 years we estimate that the minimum angular separation to be 516 $\mu$as that is within the resolving capabilities of Gaia. We searched a 0.1 deg$^{2}$ overlapping area with Gaia17bpp through both Gaia DR2 and DR3 catalogs to look for any detected sources that might be associated with this system. We identified two faint sources (Gaia DR3 1636148068921376768 with G$_{mag}$=21.1, and Gaia DR3 1636148068921376768 with G$_{mag}$=20.5) within this cone search, however, none of them seemed to be associated with the Gaia17bpp due to the lack of astrometric solutions or large parallax. On the other hand, the non-detection of the secondary star was not a surprise. First, we are limited to shallow NUV detection due to the high line of sight extinction and distance of Gaia17bpp. Second, if the secondary is surrounded by an extended disk, then it is possible the companion star is hiding within such like the $\epsilon$-Aur B5V stellar companion \citep{2012JAVSO..40..618S}. We constructed a simple toy model of a blackbody mixture model that incorporates the blackbody emission flux from the primary, and a hypothetical secondary at the same distance, including the effects of Galactic attenuation. In Figure \ref{fig:blackbodyscenario} we vary the temperature and radius of the hypothetical secondary star and integrate the blackbody mixture model to estimate the approximate SDSS-u AB magnitude. We include in Figure \ref{fig:blackbodyscenario} the derived SDSS-u $\sim$22 mag from our images as a dashed line. Based on our limiting magnitude, we cannot exclude the presence of a main sequence, white dwarf, or neutron star companion based on the observations alone. Further deep observations in the NUV and FUV will be required. 

In Figure \ref{fig:comp_dips} we compare the relative photometric depth of Gaia17bpp with other giant dimming stars from obscured by occulting disks, sorted from the shortest to the longest duration dimming events. It is evident that the Gaia17bpp system has by far the longest and deepest dimming event found amongst these candidates. It is possible that the variation in light curve properties we see here is due to the geometric configurations of such disk companions. Based on the archival photometry alone it is unclear if Gaia17bpp is periodic. Based on the archival detections from DASCH, and other surveys, we find it likely that no other major dimming event has occurred. Given the detections we have made from DASCH and POSS, we can assume a lower limit to the orbital period of at least 66 years. This would be in agreement with our occulting disk model which would suggest a very long orbital period given the very low transverse velocity. \\

\section{Conclusion}\label{sec:conclusion}
In conclusion, we report the serendipitous discovery of Gaia17bpp/2MASS J19372316+1759029 an anomalous star that exhibited a single, $\sim$7-year long, and 4.5 magnitudes deep dimming event. Using a collated dataset of multi-band light curves across several surveys and wavelength regimes, we find that this unique system closely resembles the variability seen from evolved giant stars transited by companion stars with opaque disks. We constrain the primary star of Gaia17bpp through a Bayesian amalgamation of stellar SED models. We conclude that the primary star is likely a 58$R_{\odot}$ M0-III giant star favoring a low-mass primary star that is consistent with the radius and mass profile of an ordinary giant M giant star. The long and complex nature of the dimming event is not fully understood. We find evidence that the WISE IR light curves are shallower, asymmetric, and slightly different shapes from their optical counterparts. The color evolution of the event is also perplexing, with a blue excess at the bottom of the eclipse in the infrared colors, and with others in the optical. We suspect that the main culprit of the dimming event is linked to an emerging population of rare binary systems with a companion enshrouded in a debris disk, such as the case of $\epsilon$-Aur. Finally, We performed a simple tilted-disk eclipse model of the W$_{1}$ infrared light curve. We found that the disk would need to be inclined by 178 deg, and slow-moving with an approximate radius of 1.4 AU. Based on the currently available data, we cannot identify a secondary companion star. 

Slow and photometrically deep dimming stellar systems will become even more relevant in the near future. The forthcoming Vera C. Rubin Observatory Legacy Survey of Space and Time (LSST; \cite{2019ApJ...873..111I}) promises a survey of long duration and unprecedented photometric depth. LSST will contain both a long-time baseline and unprecedented photometric depth in $ugrizy$ that will lead to the discovery of many more such extraordinary eclipsing systems. 

\section{Acknowledgments}
We thank Christina Hedges and Tobin Wainer for their useful conversation regarding the TESS photometry in crowded Galactic fields. We also thank Robert Stencel for useful comments concerning Epsilon Aurigae. Based on data from CMC15 Data Access Service at CAB (CSIC-INTA). This project was partially funded by the Department of Astronomy Jacobsen Fund at the University of Washington. This publication makes use of data products from the Wide-field Infrared Survey Explorer and NASA/IPAC Infrared Science Archive, which is a joint project of the University of California, Los Angeles, and the Jet Propulsion Laboratory/California Institute of Technology, funded by the National Aeronautics and Space Administration. 

\software{\texttt{Astropy} \citep{astropy:2013, astropy:2018, astropy:2022}, \texttt{IPython} \citep{PER-GRA:2007},\texttt{Matplotlib} \citep{Hunter:2007}, \texttt{NumPy} \citep{harris2020array}, \texttt{SciPy} \citep{2020SciPy-NMeth}}

\bibliography{sample631}{}
\bibliographystyle{aasjournal}

\end{document}